\documentstyle[12pt,epsfig]{article}

\newcommand{\bq}{\begin{equation}}
\newcommand{\eq}{\end{equation}}
\newcommand{\bqn}{\begin{eqnarray}}
\newcommand{\eqn}{\end{eqnarray}}
\newcommand{\nb}{\nonumber}
\newcommand{\lb}{\label}
 
\begin{document}
 
\baselineskip 0.65cm

\title{Thick de Sitter 3-Branes, Dynamic Black Holes and Localization of
Gravity} 
\author{Anzhong Wang\\
\small Department of Physics and Astronomy, Brigham Young University, Provo, 
Utah, 
84602\\ 
\small and\\
\small Departamento de F\'{\i}sica Te\'orica, Universidade do 
Estado do Rio de Janeiro\\
\small
Rua Sao Francisco Xavier 524, Maracan\~a, CEP. 20550-013, Rio de Janeiro, RJ, 
Brazil}
 
\date{\today}
\maketitle
 
\begin{abstract}

The embedding of a thick de Sitter 3-brane into a five-dimensional  bulk is 
studied, assuming  a scalar field with potential is present in the bulk.  A 
class of solutions is found in  closed form  that can represent 
a thick de Sitter 3-brane interpolating either between two dynamical black 
holes with  a $R \times S_{4}$ topology or between  two Rindler-like 
spacetimes with a $R_{2}\times S_{3}$ topology. 
The gravitational field is localized in a small region  near the
center of the 3-brane. The analysis of graviton fluctuations
shows that a zero mode exists and separates itself from
a   set of continuous modes by a mass gap. The existence of such a mass gap 
is shown to be universal. The scalar perturbations are also studied and shown 
to be stable.

\end{abstract} 




\vspace{1.cm}

\newpage

\section{Introduction}


The idea that our universe is embedded in a higher 
dimensional world has
received a great deal of renewed attention over the last couple of years (see, 
for example, \cite{SF01} and references therein). 
This interest is motivated  by the possibility of 
resolving the hierarchy problem, namely the large  difference in magnitudes 
between the Planck and electroweak scales
\cite{ADD98,RS99}, in addition to possibly  solving 
the long-standing cosmological constant problem  \cite{ADKS00}.  
According to this  scenario,  Standard  Model physics is confined to a three 
(spatial) dimensional hypersurface (often referred to as a 3-brane) in a 
larger dimensional space. While most physics is on the 3-brane, gravity 
propagates in the whole bulk spacetime. In previous considerations of such 
models, it was shown that  the bulk propagation of gravity is in
contradiction with the observational fact  that four-dimensional
gravity satisfies an inverse-square Newtonian law. However, in a model
proposed by Randall and Sundrum   (RS) \cite{RS99}, this problem was 
solved by relaxing one of the commonly used assumptions that our 
four-dimensional universe is independent of the coordinates defining the extra 
dimensions. When one does this, one can show that (even when the extra 
dimensions are infinitely large) gravity can be localized near the 3-brane, 
and Newtonian gravity can be restored at long distances. 
In particular, RS showed that a single massless graviton can be localized on 
the brane. This mode is responsible for producing 4D
gravity on the 3-brane, while additional massive modes only introduce small
corrections to the Newtonian law. 

Considering our four-dimensional universe as an infinitely thin
3-brane is an idealization, and in more realistic models the thickness of the
brane should be taken into account. It is for this reason that various thick
3-brane models   were considered \cite{DFGK00,KKS02}.  However,  these models 
tend to have  either naked singularities appearing at a finite distance
from the center of the brane or the scalar field has an  unusual potential.
While the nature of these singularities  is still unclear \cite{DFGK00}, one 
might like to  try to avoid them by starting with a
potential induced from some high energy theory, such as Superstring or
Supergravity. However, once a scalar field appears in the
bulk, curved four-dimensional spacetimes, rather than Minkowski,  can appear.  

In this paper, we consider the embedding of a thick de Sitter  3-brane
in a five-dimensional spacetime described by the metric,
\bq
\lb{eq0}
ds^{2} = e^{2A(z)}\left\{dt^{2} - e^{2\alpha t}d{\bf x}^{2} -
dz^{2}\right\},
\eq
where $z$ denotes the conformal coordinate of the extra dimension with 
 $z \in (-\infty, \infty)$, and $x^{\mu} = \{t, \; {\bf x}\} \; (\mu = 0, 1,
2, 3)$ are the usual four-dimensional Minkowskian coordinates. The potential
of the scalar field to be considered here is assumed to
take the form,   
\bq
\lb{eq1}
V(\phi) = V_{0} \cos^{2(1-n)}\left(\frac{\phi}{\phi_{0}}\right),
\eq
where $V_{0}$ and $n$ are arbitrary  constants, subject to $ 0 < n < 1$, and
$\phi_{0} \equiv  [3n(1 - n)]^{1/2}$. This form for the potential is quite 
similar
to that proposed in \cite{HSF89} for weakly interacting pseudo-Goldstone
bosons. 

We shall show that,  when the scalar field is the only source to the
five-dimensional Einstein field equations $R_{MN} - R\gamma_{MN}/2 = 
T^{\phi}_{MN}$,
thick brane solutions exist. Note that in this paper we shall choose
units such that $M^{3} = 1/4$, where $M$ is the five dimensional Planck
scale. In these thick brane  models, the naked  singularities which often 
appear in other brane world models are replaced 
by event horizons. The extension of the spacetime
beyond these horizons gives rise to either dynamical back holes with a  $R
\times S_{4}$ topology or Rindler-like spacetimes with a $R_{2}\times S_{3}$ 
topology. This is very much in the same spirit as the so-called
asymmetrically warped model proposed in \cite{CEG01}. However, a fundamental
difference between their model and ours is that in \cite{CEG01}  the induced
energy-momentum tensor (EMT) on the brane violates all three energy
conditions, while in the present case the weak and dominant (but  not the 
strong) energy conditions are satisfied in the whole bulk. We also find 
that the spectrum of graviton fluctuations  has a mass gap that separates the 
massless mode from a set of continuous  modes. The existence of such a 
mass gap is shown to be universal. In addition, we also study scalar 
perturbations  and show   that all the corresponding modes are stable.

\section{Solutions of Thick de Sitter 3-Branes and Dynamical Black Holes}


To show explicitly the above claims, let us 
start with 
the Einstein field equations and the corresponding Klein-Gordon equation
$\phi_{;MN}\gamma^{MN} = V'(\phi)$. It can be shown
that   there are now only two independent equations \cite{DFGK00},    
\bqn
\lb{eq2.a}
A'' + {A'}^{2} - {\alpha}^{2} &=& - \frac{1}{6}\left({\phi'}^{2}
+ 2 e^{2A}V(\phi)\right),\\
\lb{eq2.b}
{A'}^{2} - {\alpha}^{2} &=&  \frac{1}{12}\left({\phi'}^{2}
- 2 e^{2A}V(\phi)\right), 
\eqn
where the semicolon denotes the covariant derivative with respect to the bulk 
metric $\gamma_{MN}$, and a prime denotes ordinary differentiation with 
respect to $z$. Integrating these equations, we find the   solutions,     
\bq
\lb{eq3}
A = - n\ln\left[\cosh(\beta z)\right],\;
\phi = \phi_{0}\; {\sin^{-1}}\left[\tanh(\beta z)\right],
\eq
where, in terms of $V_{0}$ and $n$, the constants $\alpha$ and $\beta$ are
given via the relations $\alpha^{2} = n^{2}\beta^{2} = 2nV_{0}/[3(1 + 3n)]$.
As is well-known, the EMT for a scalar field is energetically equivalent to
an anisotropic fluid, $T_{MN} = \rho (\gamma_{MN} + z_{M}z_{N}) + p 
z_{M}z_{N}$, where $z_{M} = e^{A}\delta^{z}_{M}$ and  
\bqn
\lb{eq4}
\rho &\equiv& \frac{1}{2}\left[(\nabla \phi)^{2} + 2 V(\phi)\right] 
= \frac{2(1+n)V_{0}}{(1 + 3n)\cosh^{2(1-n)}(\beta z)},\nb\\
p &\equiv& \frac{1}{2}\left[(\nabla \phi)^{2} - 2 V(\phi)\right] 
= - \frac{4nV_{0}}{(1 + 3n)\cosh^{2(1-n)}(\beta z)},
\eqn
which shows clearly that the   thick 3-brane is localized in the
region $|z| \approx 0$, and the corresponding EMT satisfies the weak and
dominant energy conditions, but not the strong one.  
On the other hand, restoring the units, one can show that the reduced
four-dimensional Planck mass on the brane, $M_{{\em Pl}}$, is  given  by  
\bq
\lb{eq5}
\frac{M_{{\em Pl}}^{2}}{M^{3}} = \int^{\infty}_{-\infty}{e^{3A(z)} dz} 
= \frac{2^{1+3n}\Gamma^{2}(1 + 3n/2)}
{3n\beta \Gamma(1 + 3n)},
\eq
 where $\Gamma(m)$ denotes the usual gamma function. Note that this is finite 
for any $n \in (0, 1)$.

To  study  the above solutions further, let us first note that all the scalars 
built from the Riemann tensor are finite in the whole bulk, $
-\infty < x^{M} < \infty,\; (M = 0, 1, ..., 4)$.  Thus, the spacetime
described  by the above solutions are free of scalar singularities
\cite{ES77}. However, for the cases where $1/2 < n < 1$, the hypersurfaces
$|z| = \infty$ actually represent non-scalar spacetime singularities, as the
tidal forces experienced by a freely falling observer become unbounded there.
To show this explicitly,  let us consider the time-like geodesics
perpendicular to the 3-brane, which can be shown to allow the first integral
\bq
\lb{eq5a}
E^{M}_{(0)} \equiv  \dot{t}\delta^{M}_{t} + \dot{z}\delta^{M}_{z} =
\cosh^{2n}(\beta z)\left\{E\delta^{M}_{t} \pm \left[E^{2} - \cosh^{-2n}(\beta
z)\right]^{1/2}\delta^{M}_{z}\right\},
\eq
 where $E$ denotes the total energy of
the test particles and $\tau$ their proper time. From $E^{M}_{(0)}$ we can
construct  four other space-like unit vectors that are parallelly transported
along the geodesics, 
\bq
\lb{eq5b}
E^{M}_{(4)} =
 \dot{z}\delta^{M}_{t} + \dot{t}\delta^{M}_{z},\; E^{M}_{(i)} = e^{\alpha
t}\delta^{M}_{i},\; (i = 1, 2, 3),
\eq
which satisfy the relations, 
\bq
\lb{eq5c}
E^{M}_{(A); N} E^{N}_{(B)}  = 0,\;\;\;
 E^{M}_{(A)}E^{N}_{(B)}\gamma_{MN} = \eta_{AB}.
\eq
Projecting the Riemann tensor into this orthogonal frame, we find that some of 
its components, which represent the tidal forces, become unbounded at $|z| =
\infty$ for $1/2 < n < 1$. As one example, consider
\bq
\lb{eq6}
R_{(0)(2)(0)(2)} = n\beta^{2}\cosh^{2(2n-1)}(\beta z)
 \left\{(1-n)E^{2} - \cosh^{-2n}(\beta z)\right\}.
\eq
It is interesting to note that the distortion, which is equal to twice the
integral of the tidal forces with respect to $d\tau$, is finite as the
hypersurfaces $|z| = \infty$ approach. In this sense these singularities are 
weak \cite{Ori93}. When $0 < n \le 1/2$, these surfaces
represent horizons. This can be seen, for example, by calculating the proper
distance in the perpendicular direction to the brane, which is finite.  Thus,
in this latter case the spacetime needs to be extended beyond these surfaces.
Because of the reflection symmetry of the spacetime, it is sufficient
 to consider the extension across the surface $z = \infty$.  To this end,
let us first consider the coordinate transformations
\bq
\lb{eq7}
v = \alpha^{-1} e^{-\alpha (t + z)},\;\;\; 
u = \alpha^{-1} e^{\alpha (t - z)},
\eq
in terms of which the above solutions take the form
\bqn
\lb{eq8}
ds^{2} &=& 2^{2n}\left[1 + \left(- \alpha^{2} u v\right)^{1/n}\right]^{-2n}
\left[ dudv - (\alpha v)^{2}d{\bf x}^{2}\right],\nb\\
\phi &=& \phi_{0}\; {\sin^{-1}}\left[\frac{1 - \left(- \alpha^{2} u
v\right)^{1/n}}{1 + \left(- \alpha^{2} u v\right)^{1/n}}\right].
\eqn
The corresponding Kretschmann scalar is given by 
\bq
\lb{eq9}
{\cal{R}}  = 8\alpha^{2}\left(3\alpha^{2} + 2\beta^{2}\right)
\left[\frac{2\left(- \alpha^{2} u v\right)^{1/2n}}
{1 + \left(- \alpha^{2} uv\right)^{1/n}}\right]^{4(1-n)}.
\eq
The coordinate transformations (\ref{eq7}) hold only in the region $v \ge 0,\;
u \le 0$. Thus, in the whole $uv$-plane  we  obtain, in general, three
extended regions, $I',\; II$ and $II'$, as shown by Figs. 1 and 2. The 
singular properties of
the spacetime in these extended regions depend on the free parameter $n$.
In particular, when $n^{-1}$ is an integer, the extension is analytic
across the hypersurfaces $u = 0$ and $v = 0$. Otherwise, it is at 
best maximal. As a matter of fact, when $n = (2l)/(2m + 1)$, where $l$
and $m$ are positive integers, the metric in the extended regions becomes
complex, which indicates that the above extension  is not even applicable  to 
this case. However, in the rest of this paper, we shall consider only the case 
where the extension is analytic, that is, 
$n^{-1}$ is an integer. This can, in turn, be divided into two subcases,
$\; n = (2 l + 1)^{-1}$ and $\; n = (2 l)^{-1}$.

Let us first consider the case  $n = (2 l + 1)^{-1}$. Then, from 
Eqs.(\ref{eq8}) and (\ref{eq9}) we can see that the spacetime becomes singular 
on the hypersurfaces $uv = \alpha^{-2}$ in the regions $II$ and $II'$, which 
are represented by the horizontal lines $\overline{AB}$ and $\overline{CD}$ in 
Fig. 1. The
center of the 3-brane is the vertical line $\overline{bd}$, while the line
$\overline{ac}$ represents an identical 3-brane  in the extended region $I'$. 
In
the right (left) hand side of the vertical line $\overline{bd}\; 
(\overline{ac})$, we
have $z \le 0$, the extension of which across the hypersurface $z = -\infty$
is identical to the one across the surface $z = \infty$. Thus, a geodesically
maximal spacetime in this case  consists of infinite 3-branes in the
horizontal direction, as shown in Fig.1.

\begin{figure}[htbp]
\begin{center}
\leavevmode
    \epsfig{file=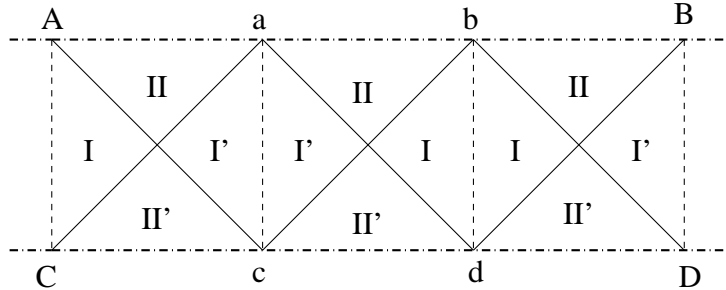,width=0.7\textwidth,angle=0}
\caption{The Penrose diagram for  the case $n = (2l +
1)^{-1}$. }    
 \label{fig1}
 \end{center}
\end{figure}


When $\; n = (2 l)^{-1}$, from Eqs.(\ref{eq8}) and
(\ref{eq9}) we can see that the spacetime is free of any kind of singularities
in all the $uv$-plane. Thus, the geodesically complete spacetime now consists
of infinite diamonds, as shown by Fig.2. The spacetime structure in each 
diamond is quite similar to that of Rindler spacetime. 

\begin{figure}[htbp]
\begin{center}
\leavevmode
    \epsfig{file=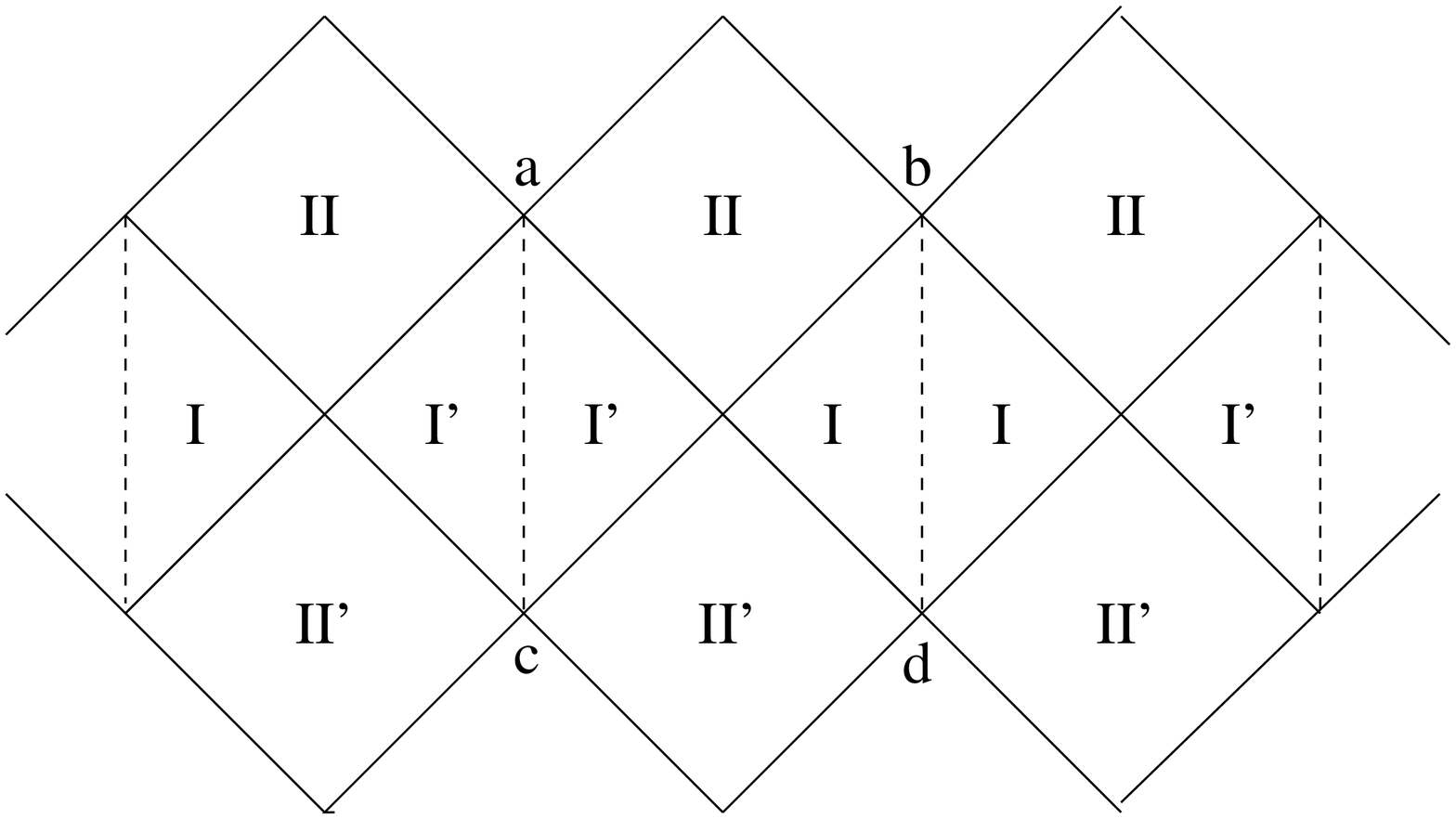,width=0.7\textwidth,angle=0}
\caption{The Penrose diagram for  the case 
$n = (2l)^{-1}$. }
\label{fig2}
 \end{center}
\end{figure}


It is remarkable to note that the requirement that the extension be analytic
automatically restricts the parameter $n$ to the range $0 < n \le 1/2$, namely 
for that range
in which we have shown that the surfaces $|z| = \infty$ represent event
horizons.

If we further introduce the coordinates,
\bqn
\lb{eq10} 
T &=& \frac{1}{2}\left[ (u + v) + \alpha^{2} v {\bf x}^{2}\right],\nb\\
Z &=& \frac{1}{2}\left[ (u - v) + \alpha^{2} v {\bf x}^{2}\right],\nb\\
{\bf X}  &=&   \alpha^{2} v {\bf x},
\eqn
we find that the above solutions can be written in the form,  
\bqn
\lb{eq8a}
ds^{2} &=& 2^{2n}\left\{1 + \left[\alpha^{2}(R^{2} - T^{2})\right]^{1/n}
\right\}^{-2n}\left(dT^{2} - d{R}^{2} -
R^{2}d\Omega^{2}_{3}\right),\nb\\   
\phi &=& \phi_{0}\; {\sin^{-1}}\left\{\frac{1 - \left[\alpha^{2}(R^{2} - 
T^{2})\right]^{1/n}}{1 +
\left[\alpha^{2}(R^{2} - T^{2})\right]^{1/n}}\right\}, 
\eqn
 where 
$d\Omega^{2}_{3}\left[\equiv d\psi^{2} + \sin^{2}\psi(d\theta^{2} 
+ \sin^{2}\theta d\varphi^{2})\right]$ denotes the metric on the unit 
three-sphere with intrinsic coordinates $\theta,\; \varphi$ and $\psi$, 
related to the Minkowskian coordinates ${\bf X}$ and $ Z$ in the usual 
way. In particular, from
Eq.(\ref{eq10}) we find that
\bq
\lb{eq11}
R^{2} - T^{2} = - uv,
\eq
from which we can see that the hypersurface $z = 0$ or $uv = - \alpha^{-2}$
is a bubble with  constant acceleration. When $n = (2l + 1)^{-1}$, a 
spacetime singularity  develops on the hypersurfaces $R^{2} - T^{2} = - 
\alpha^{-2}$ in regions $II$ and $II'$, where the geometric radius vanishes.  
Thus,  the spacetime in this case actually has   $R\times S_{4}$ 
topology. When $n = (2l)^{-1}$, no such singularities are formed, and the 
spacetime extends to the whole range $R\in [0, \; \infty)$. Consequently, the 
spacetime has  $R_{2}\times S_{3}$ topology.

\section{Gravitational and Scalar Perturbations}


The analysis of the metric fluctuations in general 
is complicated because of their coupling to the the scalar field fluctuations. 
This is particularly true when the four-dimensional spacetime is curved as in 
the present case. Following \cite{GS99},  let us first consider the 
gravitational perturbations,
\bq
\lb{eq12}
ds^{2} = e^{2A(z)}\left[(g_{\mu\nu} + h_{\mu\nu})dx^{\mu}dx^{\nu} -
dz^{2}\right],
\eq
where $g_{\mu\nu}$ denotes the 4D background metric, and $h_{\mu\nu}$ the
metric perturbations, satisfying the transverse-traceless (TT) condition,
\bq
\lb{eq12a}
h^{\lambda}_{\lambda} = 0 = h_{\mu\nu;\lambda}g^{\nu\lambda}.
\eq
Then, it can be shown that the equation for ${h}_{\mu\nu}$  
is given by \cite{KKS02}
\bq
\lb{eq13}
h''_{\mu\nu} + 3 A'h'_{\mu\nu} - \Box h_{\mu\nu} 
- 2 \alpha^{2} h_{\mu\nu} = 0, 
\eq
where $\Box  \equiv g^{\alpha\beta}\nabla_{\alpha}\nabla_{\beta}$ and 
$\nabla$ denotes the covariant derivative with respect to the 4D metric 
$g_{\mu\nu}$. Since we are looking for a mode that corresponds to a 4D
graviton, let us define the mass of such a spin two excitation by \footnote{I 
would like to thank A. Karch for pointing out an error in an earlier 
definition for the mass in the de Sitter background.}
\bq
\lb{eq13a0}
\Box h_{\mu\nu} + 2\alpha^{2} h_{\mu\nu} = - m^{2}h_{\mu\nu}.
\eq
Then, following the standard procedure we introduce the polarization tensor 
$\epsilon_{\mu\nu}(x^{\alpha})$ via the relations
\bq
\lb{eq13a}
h_{\mu\nu}(x^{\alpha}, z) = e^{-3A/2}\epsilon_{\mu\nu}(x^{\alpha})\Psi(z),
\eq
where $\epsilon_{\mu\nu}$ satisfies the TT condition (\ref{eq12a}). Inserting 
Eqs.(\ref{eq13a0}) and (\ref{eq13a}) into Eq.(\ref{eq13}),  we find that the 
equation for $\Psi(z)$
\bq
\lb{eq13b}
\left(- \partial^{2}_{z} + V_{QM}(z)\right)\Psi(z) = m^{2} \Psi(z),
\eq
with   the effective potential $V_{QM}(z)$
given by
\bq
\lb{eq14}
V_{QM}(z) \equiv \frac{9}{4}{A'}^{2} + \frac{3}{2}A''
= \frac{3n\beta^{2}}{4}\left(3n 
- \frac{3n + 2}{\cosh^{2}(\beta z)}\right).
\eq
When $ m = 0$, Eq.(\ref{eq13b}) has the solution
\bq
\lb{eq14b}
\Psi_{0}(z) = e^{3A(z)/2} = \cosh^{-3n/2}(\beta z).
\eq
Since $0 < n \le 1/2$, it can be seen that this zero mode is normalizable.
Following a similar argument as that in \cite{RS99,DFGK00}, one can show that
this zero mode will give rise to  4D gravity on the 3-brane.

On the other hand, from Eq.(\ref{eq14}) we can see that $V_{QM}(z) \rightarrow 
9\alpha^{2}/4$, as $|z| \rightarrow \infty$. Thus, similar to thin de Sitter 
3-branes \cite{GS99},  there is also  a mass gap in the present case. It is 
interesting to note that the existence of such a mass gap is actually a 
universal property  for de Sitter 
3-branes.  To see this, let us first notice that for any given scalar 
potential $V(\phi)$, the effective potential $V_{QM}$ can be written as
\bq
\lb{eq14a}
V_{QM}(z) = \frac{9}{4}\alpha^{2} - \frac{1}{8}e^{2A}\left(4\rho - p\right),
\eq
where $\rho$ and $p$ are defined by Eq.(\ref{eq4}). Thus, as long as the 
gravitational  field is localized in the region $|z| \sim 0$, 
the second term at the right-hand side of Eq.(\ref{eq14a})  goes to zero as 
$|z| \rightarrow \infty$, while the effective potential $V_{QM}$  goes to 
$9\alpha^{2}/4$.

To study the above problem further, let us introduce a new variable $x$ by $x 
= \tanh(\beta z)$, then we find that Eq.(\ref{eq13b}) becomes the standard 
Legendre differential equation,
\bq
\lb{eq15}
\left\{(1 - x^{2})\partial^{2}_{x}  - 2x\partial_{x} + \left[\nu(\nu + 1) -
\frac{\mu^{2}}{1 - x^{2}}\right]\right\}\Psi(x) = 0,
\eq
with 
\bq
\lb{eq15a}
\nu \equiv \frac{3}{2}n,\;\;\;\; \mu^{2} \equiv - 
\frac{1}{\beta^{2}}\left(m^{2} - 
\frac{9}{4}\alpha^{2}\right).
\eq
The general solution of the above equation is given by
\bq
\lb{eq16}
\Psi(x) = C_{1} P^{\mu}_{\nu}(x) + C_{2}Q^{\mu}_{\nu}(x),
\eq
where $P^{\mu}_{\nu}(x)$ and $Q^{\mu}_{\nu}(x)$ denote the associated Legendre
functions of the first and second kinds, respectively. When $\mu$ is real, or
$m^{2} < 9\alpha^{2}/4$, both $P^{\mu}_{\nu}(x)$ and $Q^{\mu}_{\nu}(x)$ are 
singular at $|x| = 1\;
({\rm  or }\; |z| = \infty)$. Thus, the regularity conditions at $|x| = 1$ for
the perturbations exclude the case $m^{2} < 9\alpha^{2}/4$.  When $\mu = 0$ 
or $m^{2}  = 9\alpha^{2}/4$, the functions $P^{\mu}_{\nu}(x)$
and $Q^{\mu}_{\nu}(x)$ reduce to the Legendre polynomials of the first and 
second kinds, $P_{\nu}(x)$ and $Q_{\nu}(x)$, respectively, and the latter is
still singular at  $|x| = 1$. Thus, the regularity conditions at $|x| = 1$
force $C_{2} = 0$, and  we obtain $\Psi_{0}(x) = C_{1} P_{\nu}(x)$,  which is 
always finite and normalizable.  The case where
$m^{2} > 9\alpha^{2}/4$ corresponds to a continuous spectrum of eigenfunctions 
that asymptote to plane waves in the limit $|z| \rightarrow \infty$. Similarly,
one can show that these continuous modes will produce  small corrections to the
Newtonian law as in the flat 4D case \cite{RS99,DFGK00}.

Now let us consider the scalar perturbations given by \cite{KKS02}
\bq
\lb{eq17}
ds^{2} = e^{2A}\left\{(1 + 2\varphi)dz^{2} + (1 + 
2\psi)g_{\mu\nu}dx^{\mu}dx^{\nu}\right\}.
\eq
As shown in \cite{KKS02}, the corresponding linearized 5D Einstein-scalar 
equations become
\bqn
\lb{eq18a}
& & \delta\phi = \frac{3}{\phi'}\left(\varphi A' - \psi'\right),\;\;\;
\varphi = - 2\psi,\\
\lb{eq18b}
& & \left(- \partial^{2}_{z} + V_{eff.}(z)\right)\chi(x^{\alpha}, z) = \Box 
\chi(x^{\alpha}, z),
\eqn
where $\delta\phi$ denotes perturbations of the scalar field, and 
$\chi(x^{\alpha}, z)$ and $V_{eff.}(z)$ are defined as
\bqn
\lb{eq19}
\chi(x^{\alpha}, z) &\equiv& \frac{1}{\phi'(z)}e^{3A/2}\psi(x^{\alpha}, 
z),\nb\\
V_{eff.}(z) &\equiv& - \frac{5}{2}A'' + \frac{9}{4}{A'}^{2} 
+ A'\frac{\phi''}{\phi'} - \frac{\phi'''}{\phi'}
+ 2\left(\frac{\phi''}{\phi'}\right)^{2} - 6 \alpha^{2}\nb\\
&=& \frac{\beta^{2}}{4\cosh^{2}(\beta z)}\left[2\left(2 + 5n - 12n^{2}\right)
\right.\nb\\
& & \;\; \left. + \left(4 + 4n - 15n^{2}\right)\sin^{2}(\beta z)\right].
\eqn
Because $V_{eff.}(z)$ is always positive for $0 < n \le 1/2$, following the 
arguments given in \cite{KKS02}, it can be shown that in the present case all 
the corresponding perturbation modes are stable. This is to be contrasted with 
the case of thin 3-branes \cite{GS01}, but it is a similar result to that  for 
thick 3-branes 
\cite{KKS02}.

\section{Conclusions}


In this paper, we have considered the embedding of thick de 
Sitter 3-branes in a 5D bulk in which a scalar field with potential given by
Eq.(\ref{eq1}) is assumed to be present. A class of solutions   has been
obtained in closed form, some of which represent a thick de Sitter 3-brane
interpolating between two dynamical black holes with a $R \times S_{4}$ 
topology, and some of which represent such a 3-brane
interpolating between two Rindler-like spacetimes with a $R_{2} \times S_{3}$ 
topology. The thick brane is localized in the region where $|z| \approx 0$. The
analysis of graviton fluctuations shows that the spectrum of perturbations
consists of a zero mode and  a set of continuous modes. The massless mode is 
separated by a mass gap from the continuous modes. The existence of such a 
mass gap has been shown to be universal for all such de Sitter 3-branes. 
The scalar perturbations of the solutions have also been studied and 
found to be stable.

Finally, we note that by making the replacement, $\left(t,\; x^{1}, 
\alpha\right) \rightarrow \left(i x, 
\; i t,\; - i \alpha\right)$ in the solutions given by (\ref{eq0}) - 
(\ref{eq3}), we 
can obtain solutions that represent thick anti-de Sitter 3-branes, given by
\bq
\lb{eq20}
ds^{2} = e^{2A(z)}\left\{dz^{2} + dx^{2} + 
e^{2\alpha x}\left[-dt^{2} + \left(dx^{2}\right)^{2} 
+ \left(dx^{3}\right)^{2}\right]\right\},
\eq
with
\bqn
\lb{eq21}
V(\phi) &=& V_{0}\cosh^{2(1-n)}\left(\frac{\phi}{\phi_{0}}\right),\nb\\
A(z)&=& - n\ln\left|\cos(\beta z)\right|,\nb\\
\phi(z)&=& \phi_{0}\sinh^{-1}\left[{\rm{tg}}(\beta z)\right],\nb\\
\alpha^{2} &=& n^{2}\beta^{2} = - \frac{2nV_{0}}{3(1 + 3n)}.
\eqn
but now with $\phi_{0} \equiv \left[3n(n-1)\right]^{1/2}$. The constants 
$V_{0}$ and $n$ are again arbitrary. However, to have $\phi_{0}$  
real, we must have $n \ge 1$ or $n \le 0$. On the other hand, to have $\alpha$ 
and $\beta$ real,  $\; V_{0}$ has to be negative for
$n > 1$ or $n < - 1/3$ and positive for $-1/3 < n < 0$. Studying these   
solutions is outside the scope of this paper, but we plan to return to this 
problem on another occasion.

\section*{ACKNOWLEDGMENTS}


The author would like to thank E. W. Hirschmann and Andreas Karch  for 
valuable suggestions and dicusssions on braneworld scenarios, and the 
Department of Physics and Astronomy, BYU, for hospitality. Financial  
assistance from BYU, CNPq and FAPERJ is gratefully acknowledged.




\begin{thebibliography}{99}

\bibitem{SF01} V.A. Rubakov, Phys. Usp. {\bf 44}, 871  (2001);  
 S. F\"oste, Fortsch. Phys. {\bf 50}, 221 (2002).

 
\bibitem{ADD98}  N. Arkani-Hamed, S. Dimopoulos and G. Dvali, Phys. Lett. {\bf 
%
B429}, 263 (1998); I. Antoniadis, N. Arkani-Hamed, S. Dimopoulos and G.
Dvali, Phys. Lett., {\bf B436},  257 (1998).

\bibitem{RS99} L. Randall and  R. Sundrum, Phys. Rev. Lett. {\bf 83},
3370  (1999); {\em ibid.}, 4690  
(1999).

\bibitem{ADKS00}  N. Arkani-Hamed, S. Dimopoulos, N. Kaloper
and R. Sundrum, Nucl. Phys. {\bf B480}, 193 (2000);  S. Kachru, M.
Schulz, and E. Silverstein, Phys. Rev. {\bf D62}, 045021 (2000).

 
\bibitem{DFGK00}
O. DeWolfe, D. Z. Freedman, S. S. Gubser and A. Karch, 
Phys. Rev. {\bf D62}, 046008  (2000);
%
M. Gremm, Phys. Lett. {\bf B478}, 434  (2000); Phys. Rev. {\bf D62}, 044017  
(2000); 
%
C. Cs\'{a}ki, J. Erlich, T. J. Hollowood and Y. Shirman,  Nucl. Phys. {\bf 
B581}, 309  (2000); 
%
S. Ichinose, Class. Quantum Grav. {\bf 18}, 5239 (2001);   
%
A. Kehagias and K. Tamvakis, Phys. Lett. {\bf B504}, 38  (2001);
``{\em A Self-Tuning Solution of the Cosmological Constant Problem}," {\tt
hep-th/0011006} (2000);
%
K. Behrndt and G. Dall'Agata, Nucl. Phys. {\bf B627}, 357 (2002).

\bibitem{KKS02} S. Kobayashi, K. Koyama, and J. Soda, Phys. Rev. {\bf D65}, 
064014 (2002).


 
\bibitem{HSF89} C.T. Hill, D.N. Schramm, and J.N. Fry, Comm. Nucl. Phys. {\bf
19}, 25 (1989); G. Goetz, J. Math. Phy. {\bf 31}, 2683 (1990); M. Mukherjee,
Class. Quantum Grav. {\bf 10}, 131 (1993).

\bibitem{CEG01} C. Csaki, J. Erlich, and C. Grojean, Nucl. Phys. {\bf B604},
312 (2001).


\bibitem{FGPW99} D.Z. Freedman, S.S. Gubser, K. Pilch, and N.P. Warner, 
JHEP, {\bf 0007}, 038 (2000).

\bibitem{ES77} G.F.R. Ellis and B.G. Schmidt, Gen. Relativ. Grav. {\bf 8}, 915
(1977).

\bibitem{Ori93}  L.M. Burko and  A. Ori,  Phys. Rev. Lett. {\bf 74}, 
1064 (1995), and references therein.   

\bibitem{GS99} J. Garriga and M. Sasaki, Phys. Rev. {\bf D62}, 043523 (2000); 
N. Alonso-Alberca, P. Meessen, and T. Ortin, Phys. Lett. {\bf B482}, 400 
(2000); A. Karch and L. Randall, JHEP, {\bf 0105}, 008 (2001).

\bibitem{GS01} U. Gen and M. Sasaki, Prog. Theor. Phys. {\bf 105}, 591 (2001); 
Z. Chacko and P.J. Fox, Phys. Rev. {\bf D64}, 024015 (2001).



\end{thebibliography}
\end{document}